\documentclass[12pt,preprint]{aastex}

\usepackage{psfig}

\def\msol{\hbox{$M_\odot$}}
\def\om{\Omega_B h^2}
\def\e10{\eta_{10}}

\def\etal{{\it et al.\ }}
\def\iso#1#2{\mbox{${}^{#2}{\rm #1}$}}
\newcommand\he[1]{\iso{He}{#1}}
\newcommand\li[1]{\iso{Li}{#1}}

\def\li#1{\iso{Li}{#1}}
\def\b1#1{\iso{B}{1#1}}

\def\beq{\begin{equation}}
\def\eeq{\end{equation}}
\def\beqar{\begin{eqnarray}}
\def\eeqar{\end{eqnarray}}
\def\simlt{\lower.5ex\hbox{$\; \buildrel < \over \sim \;$}}
\def\simgt{\lower.5ex\hbox{$\; \buildrel > \over \sim \;$}}
\def\simpropto{\lower.2ex\hbox{$\; \buildrel \propto \over \sim \;$}}

\begin{document}


\title{On the Possible Sources of D/H Dispersion at High Redshift}

\author{Brian D. Fields}
\affil{Department of Astronomy, University of Illinois,
Urbana, IL 61801, USA}

\author{Keith~A.~Olive}
\affil{ Theoretical Physics
Institute, School of Physics and Astronomy, \\
University of Minnesota, Minneapolis, MN 55455 USA}

\author{Joseph Silk}
\affil{Department of Physics, University of Oxford, Keble Road, Oxford OX1 3RH,\\
and Institut d'Astrophysique, 98 bis Boulevard Arago,
Paris 75014, France}

\author{Michel Cass\'{e}}
\affil{Service d'Astrophysique, CEA, Orme des Merisiers,
91191 Gif sur Yvette, France \\
also Institut d'Astrophysique, 98 bis Boulevard Arago,
Paris 75014, France}

 \and \author{Elisabeth Vangioni-Flam}
\affil{Institut d'Astrophysique, 98 bis Boulevard Arago,
Paris 75014, France}

\begin{abstract}

\vskip-6.9in
\begin{flushright}
UMN-TH-2018/01 \\
TPI-MINN-01/34 \\
astro-ph/0107389 \\
July 2001
\end{flushright}
\vskip+5.7in

Recent observations suggest the existence of a
white dwarf population in the Galactic
halo, while others suggest that deuterium has been astrated
in systems at high redshift and low metallicity.
We propose that these observations could be signatures
of an early population of intermediate-mass stars.
Such a population requires a Population III
initial mass function different from that of the solar neighborhood, 
as perhaps also suggested
by the observed cosmic infrared background.
Also, to avoid overproduction of 
C and N, it is required that the
$Z=0$ yields of these stars have low ($\sim 10^{-3}$ solar) 
abundances 
as suggested by some recent calculations.
Under these assumptions, we present a model
which
reproduces the observed  D vs Si trend, and 
predicts a high cosmic Type Ia supernova rate,
while producing a white dwarf population that 
accounts for only $\sim 1.5\%$ of the dark halo.
This scenario can be tested by observations of
the cosmic supernova rate, and by confirmation and
further
studies of the putative white dwarf halo population.

\end{abstract}


\section{Introduction}

One of the key parameters in the standard big bang model
 is the density of ordinary baryonic matter, expressed as
a fraction of the critical density, $\Omega_B$ (often quoted in
combination with the Hubble parameter as $\om$). 
Big bang nucleosynthesis (BBN) has traditionally been the primary
theoretical tool for determining $\om$,
and has recently been complemented by
detailed measurements of the cosmic microwave background
anisotropy. 
These results are to be compared to
direct surveys of cosmic baryonic matter. 
While baryons
are readily observable in stars or stellar remnants, galaxies, 
X-ray-emitting hot gas, as well as intergalactic gas, 
the low-redshift baryonic budget
(Fukugita, Hogan, \& Peebles \cite{fhp})
is uncertain, and leaves some room for dark baryonic objects. 
Conversely,
at higher redshifts, ($z= 2$ to 3) a new estimate of $\om$ based on
the measurement of Lyman alpha forest intergalactic 
\ion{H}{1} absorption lines (and also \ion{He}{2}
lines) in the spectra of  remote quasars has been done and is in
good agreement with $\om$ derived from BBN 
(in the range $\om = 0.01$ to $0.03$, Riedeger \etal \cite{rie}, Wadsley
\etal \cite{wad}).
This is between 3 and 20 times the observed baryon density
in the luminous parts of galaxies and their presumed precursors, the
damped Lyman alpha intergalactic  neutral hydrogen clouds, amounting to 
 $\om = 0.0015$ to $0.003$ (P\'eroux \etal 2001).

At the local scale,
MACHOS, with masses of about 0.5 $M_\odot$
and detected by gravitational microlensing towards the LMC,  are possible
candidates for  some of  the dark baryons, and amount to $\simlt 20\%$ of the
dark halo (Alcock \etal 2000). Recently, old halo white dwarfs may have been detected in
sufficient numbers to amount to
$\simgt 2\%$ of the dark halo. These white dwarfs, if halo members, 
represent a large increment, by a factor $\sim 10$, relative to disk white
dwarfs, and hence would  have a profound impact on Galactic chemical
evolution. In this paper we discuss the implications of the discovery of  dark
baryons in the form of old halo white dwarfs for BBN, and in particular for
the deuterium abundance.

The standard model of BBN has enjoyed great success in its ability to 
predict the primordial abundances of the light element isotopes, D,
\he3, \he4 and \li7. Indeed, in the standard model, there remains only
one remaining unknown parameter, the baryon-to-photon ratio, $\eta =
n_B/n_\gamma$, which can be expressed as $\eta_{10} = 10^{10} \eta = 274
\om$. Roughly, for a value of $\eta_{10}$ between 1.7 and 7.4,
(corresponding to $\om$ between 0.006 and 0.027), we obtain concordance
between the BBN predictions and the observational determination of the
abundances (Nollett \& Burles 2000; 
Vangioni-Flam, Coc, \& Cass\'e 2001;
Cyburt, Fields, \& Olive
2001). In detail, however, there are small but perhaps not
insignificant differences in the determination of
$\om$ from the different isotopes.  \he4 and \li7 concordance is typically
achieved for lower $\e10$ whereas D/H is best fit at higher values.  To
be sure, not every isotope serves as an equally good baryometer. In fact,
because of the weak logarithmic dependence on $\e10$, \he4 is the worst
element to use to fix the baryon density.  While the  \li7 abundance
strongly depends on $\e10$, the theoretical uncertainties which stem from
uncertainties in the input nuclear cross-section make it difficult to
derive precise values of the baryon density. In addition, because of
the functional dependence of \li7 on $\e10$, a given observational
abundance of \li7 gives us two predictions for $\e10$. D/H, on the
other hand, is both strongly dependent on $\e10$ and is relatively free
from theoretical uncertainties.  One should bear in mind however, that
there are far  more observational determinations of \he4 and \li7 at the
present time than there are of D/H from quasar absorption systems.
Nevertheless it is the range in
D/H, reported to span up to a factor of 20,
that has been the center of much controversy over the past decade because of
the strong sensitivity to the baryon density. The reported D/H abundance
has now converged towards the lower end of this range, but recent
measurements reveal a spread by up to  a  factor of  3  that may reflect
intrinsic variations in the D/H abundance. These variations would most
plausibly arise from stellar processing of primordial gas,
or possibly from non thermal processes such as
photodisintegration by gamma rays, or spallation reactions
induced by energetic particles.

The  recent
results BOOMERanG 
yield  $\om = 0.021^{+0.004}_{-0.003}$ (Netterfield \etal 2001),
in very good agreement with the estimate from D/H. 
This is also in agreement with the recent results of DASI (Pryke
\etal 2001), who find $\om = 0.022^{+0.004}_{-0.003}$.

There have also been several new observations of D/H from quasar
absorption systems.  In addition to the earlier determinations by
Burles and Tytler (1998a,b) in two Lyman limit systems of $4.0 \pm .65$
and $(3.25 \pm 0.3) \times 10^{-5}$ and the upper limit in a third Lyman
limit system by Kirkman
\etal (1999) of 6.8 $\times 10^{-5}$, there have been three new
measurements of D/H in Damped Lyman $\alpha$ systems by O'Meara \etal
(2001): $2.5 \pm 0.25$, by D'Odorico \etal (2001): $2.25 \pm 0.65$, and
by Pettini and Bowen (2001): $1.65 \pm 0.35$ all times $10^{-5}$.  This
last value is uncertain due
to the poorly defined level of the continuum and the possible blending of
lines (P. Petitjean, private communication).
We
note that there is also a measurement of D/H in a relatively low
redshift system with significantly higher D/H (Webb \etal 1997; Tytler
\etal 1999).  This system has recently been criticized (Kirkman \etal
2001) and we will not consider it further here. 

Is there a real dispersion in D/H in these high redshift systems?  
We note two possible
trends that go in a similar direction. The data may show an inverse 
correlation of D/H abundance  with Si. This
may be an artifact of poorly determined Si abundances, or
(as yet unknown) systematics affecting the
D/H determination in high-column density (damped Lyman-$\alpha$, 
hereafter DLA) 
or low-column density 
(Lyman limit systems) absorbers.
On the other hand, if the dispersion in D/H is
real, it has profound consequences, as 
it must indicate that {\em some processing of  D/H must have
occurred even at high redshift}. 
The second trend is that the data may show 
an inverse  correlation of D/H abundance  with HI column density.
If real, this would suggest that in the high column density DLA systems, 
which are most likely to have undergone some star formation,  some processing
of  D/H must  similarly have occurred, likewise  at high redshift.

It is interesting to note that local D values measured in the interstellar
medium  also appear to show a considerable dispersion. 
Linsky \cite{linsky} summarized the observations then
available and argued for a constant value in the local ISM 
of $(1.5 \pm 0.1) \times 10^{-5}$.
Recent data (see Vidal-Madjar
2001, for a review) are consistent with this mean value,
but suggest significant variations of the D/H ratio spanning
the range $ 5 \times 10^{-6}$ to $ 4 \times 10^{-5}$.
Thus, if D shows such variability, perhaps correlated to some unexpected 
physical processes leading to its destruction, one can, in turn,
question the ability of cosmological clouds to represent the genuine
primordial abundance.

Thus, a simple average of D/H abundance determinations does not
make sense, at least without a proper enlargement of the error in the
mean due to the poor $\chi^2$ that such a mean would produce. Moreover, if
deuterium destruction has occurred, we must question the extent to which
any of these systems determine the value of $\Omega_B$. 

There is one obvious objection to the hypothesis that 
nucleosynthetic processing might have
occurred in the high-redshift systems. 
The absorbing systems have low metallicity. We now point to 
seemingly unrelated observations that indicate that primordial processing 
may have occurred, and with substantially less enrichment than is associated
with standard yields for a solar neighborhood initial mass function (IMF).

A substantial population of low
luminosity white dwarfs has recently been discovered
(Oppenheimer \etal\ 2001; Ibata \etal\ 2000). 
These objects were identified by their large proper motion,
indicating halo kinematics.
Oppenheimer \etal\ find that if the objects are a halo population
of white dwarfs, then they represent a fraction
$\ga 2.5\%$ of the dark halo mass.
There has been some question about whether these objects
do reside in the halo (Hansen 2001; Reid, Sahu and Hawley 2001),
but a recent re-analysis of the data by 
Koopmans \& Blandford \cite{kb}
also concludes that these are halo objects, and
finds a similar mass budget to that of Oppenheimer \etal
We now address the potential implications of
any  observed dispersion in D/H in high redshift quasar absorption systems
for a  relationship between
the possibility of D/H destruction in these systems by intermediate mass stars and the
ensuing production of a population of halo white dwarfs.

We summarize the importance of D as a baryometer, and
its relation to the other light element abundances, in \S2.
In \S3 we argue that the observed dispersion in high-redshift D/H,
if real,  
cannot be accounted for by
standard chemical evolution models.
Thus, an explanation of such a dispersion
will necessarily involve nonstandard 
scenarios, involving, e.g., variations in the IMF, or
outflows of material.
We search for models which can explain this dispersion
as well as the white dwarf observations. 
We find that such models can be constructed, 
assuming an initial burst of intermediate-mass star formation.
It is also necessary that the C and N yields
of zero metallicity stars are small, as suggested by some
recent models. 
The prospects for testing this model
and the implications for BBN are summarized
in \S4.

\section{Determinations of $\Omega_B$}

As discussed above, each of the light elements can be individually
used to determine $\om$ with varying accuracy.  For example, a helium
mass fraction of Y = 0.238 corresponds to $\e10 = 2.4$ or $\om = 0.009$.
However, the systematic uncertainties in the \he4 abundance make it
difficult to exclude values of Y as high as 0.25 (see for example,
Olive \& Skillman 2001) which because of the weak dependence on $\eta$
corresponds to
$\e10 = 7.4$ or
$\om = 0.027
$.  With \li7 one can do much better. A plateau value of \li7/H = 1.2
$\times 10^{-10}$ corresponds to $\e10 =  2.4$ and 3.1.  However, the
slightly higher value of \li7/H = 2 $\times
10^{-10}$ (consistent within systematic uncertainties) yields
$\e10 = 1.7$ and 4.5. 
While there are occasional attempts to further increase the \li7
abundance on the basis that some \li7 has been depleted (Vauclair \&
Charbonnel 1998; Pinsonneault \etal 1998,2001; Theado \& Vauclair 2001),
it is very hard to do so, due to recent \li6 observations
which leave little room, if any, for \li7 depletion (Fields \& Olive
1999; Vangioni-Flam \etal 1999) and the lack of dispersion in the Li data
(Bonifacio
\& Molaro 1997;  Ryan, Norris, \& Beers 1999; Ryan \etal 2000).

In contrast, the strong dependence of D/H on $\eta$ makes deuterium a
very good baryometer, provided we have an accurate determination of D/H.
The values of D/H measured in the high redshift systems however cover
the range $\e10 = 4.8$ for D/H = 4 $\times 10^{-5}$ to $\e10=8.5$ for
D/H = 1.65 $\times 10^{-5}$.  Thus if the dispersion in the D/H is real
(beyond observational error as the quoted error bars imply), then one
can question the extent that any of these determinations truly
represent the primordial D/H abundance.  

In the event that the dispersion is real, some amount of D/H must have
been destroyed, presumably by stellar processing.
Since ${\rm D/H}(t)$ is monotonically
decreasing in time, the primordial value should be greater than or
equal to the maximum of the  observed values. 

Stellar processing must involve short-lived stars of intermediate or
even larger mass.  The intermediate mass stars are the proposed white
dwarf precursors.  However, even an IMF peaked near 2-8 $\msol$
could include some massive stars above 10 $\msol$ that would explode
as SNII and produce the heavy element enriched observed in the
absorption systems. An inverse correlation of D with Si abundance is a
natural consequence. Since these enriched systems are preferentially
the more massive and advanced with respect to disk formation, as we
discuss below, they might also have higher gas column densities and
account for the additional inverse correlation of D with HI column.

\section{Destruction of D/H by galactic chemical evolution}

The high column density quasar absorption systems are generally considered 
to be protogalactic disks. In particular DLAs have the kinematics
and ionization characteristic of the interstellar medium of massive protodisks.
Lyman limit systems have a greater extent
and display higher ionization than the DLAs. Star formation 
with a normal IMF could not possibly have occurred in these systems to any
substantial extent.  Typical metallicities are 0.1 to 0.01 solar, and
this is consistent with these systems being predominantly gaseous.
However in order for there to be enough halo white dwarfs to account for
the recent observations, the primordial IMF must have been very different
from the present day IMF, and  been dominated by intermediate mass stars.
These stars would have formed during the protogalactic phase, and could
have destroyed significant amounts of primordial deuterium.

Typical stellar D/H depletion mechanisms usually predict the
production of heavy elements in closed box models.  Indeed, by the time the
metallicity reaches 0.1 solar abundance  in a standard galactic evolution model, only a
few percent of the initial D/H ratio has been affected.  This is the main
reason one is led to associate the observed D/H ratio to that of the
primordial deuterium abundance. It is conceivable, however, that an early
generation of stars was formed with masses only in the intermediate mass range
of 2-8 $\msol$. These stars would not be
significant  producers of  heavy elements (oxygen and above)
 usually
ascribed to Type II supernovae. Motivation for this suggestion comes from the
discovery by Oppenheimer \etal (2001) of a population of old white dwarfs in
the halo of the  Galaxy (see also 
Ibata \etal\ etal 2000; 
Chabrier 1999; Gates \& Gyuk 2001).

Previous studies of  white dwarf halos have examined
the consequences when
the entire halo consists of  white dwarfs. There are many objections to
this extreme hypothesis, ranging from overproduction of He, excessive halo 
Type Ia
supernova rates,  excessive  light element
production in the early (proto-)Galaxy, and
excessive C and N production. 
Constraints on the high-redshift population, and
possible observable signatures have been studied by
several groups 
(Ryu, Olive, \& Silk 1990; Fields, Freese \& Graff 2000;
and references therein).
The principal refinement we make to these ideas is to note that 
the old white dwarf halo population is sub-dominant,
possibly amounting to as little as 2 percent of the dark halo.\footnote{
It has been argued (Reid \etal 2001)) that the detected white dwarfs
might be a thick disk population.  If this is the case,
the total mass of white dwarfs is difficult to determine
precisely but is considerably smaller and would thus
not have important consequences for deuterium.}
Nevertheless, the effects on galactic chemical evolution are profound.

To see this, let consider the 
mass budget for deuterium destruction.  
In order to achieve noticeable D depletion, one must
process a significant amount of material though an early generation of stars. 
A typical galaxy may have $M_B = 10^{11}
M_\odot$  of visible baryons, mostly in the form of stars, and about
$10^{12} M_\odot$ total, the balance being dark matter. 
We point out that a white dwarf halo mass fraction of even 2\% can have 
significant consequences for D destruction in protodisk and disk  gas.
This yields about $2\times 10^{10}
M_\odot$ in white dwarfs, and hence $\sim  10^{11}
M_\odot$, comparable to the present disk mass, in precursor stars. 
All $8\times 10^{10}
M_\odot$ of the ejecta is totally astrated.
Hence present-day astration factors of 2 or more  are easily achievable  with a white 
dwarf halo mass fraction of  1\% or less, in an interstellar medium containing some
$6\times 10^{9}
M_\odot$ in atomic and molecular gas.

The requirements of D astration and white dwarf production,
all with little heavy element production, lead to
very strong constraints on the required IMF.
To avoid heavy element production through Type II supernovae, we must demand
that the IMF is restricted to the 
2-8 $M_\odot$ range (Adams \& Laughlin \cite{al}).  
At the $m < 2 M_\odot$ end, the constraint is that too
many stars would be luminous today or would have been seen as a huge
population of red giants at $z \sim 3$,
as $t \simeq 2$ Gyr at this redshift
in an $\Omega_{\rm m} = 0.3$ and $\Omega_\Lambda = 0.7$
universe.
On the $m>8M_\odot$ end, the
problem is that we do not  want to make a lot of metals (oxygen and up)
from supernovae. This constraint is not as strong, and we 
need 
to add a few supernovae to get some metal production for the 
absorption systems. 

A new result which independently demands an
altered IMF at high redshift comes from the 
far IR background.
Models of the FIRB (by ULIGs at high z)  overproduce 
K light today (ie too many low mass stars) if a standard IMF is adopted
(Cole \etal 2001).
An IMF peaked at intermediate masses
offers an attractive  solution: it avoids the low mass
stars and overproduction of heavy elements
but gives the light. The FIRB requires $\nu \, i_\nu$
to be about equal to what is produced with a standard IMF
for the optical/near-IR background, i.e., about 
$10-15 \ {\rm nW/m^2}$
(and to be produced in dust-shrouded ULIGS;
see, e.g., Hauser and Dwek 2001 for a recent review). The white dwarfs that the
ULIGS (which are known to be merger-induced bursts of star formation) 
produce would then (possibly after another merger or if they formed in
outflow-boosted star formation) be the halo white
dwarfs. The kinematics are a
weak point but the moral is striking:
a peaked IMF may be needed  for completely different reasons.

The obvious objection is that the hypothesised precursor stars will
overproduce C and possibly N that would be observable in Galactic
or extragalactic material (e.g., Fields, Freese, \& Graff 2000).
This is a serious issue due to the nature of the observed
C and N abundances in primitive systems.
For our purposes, the
most constraining C and N values are those in the same QSO absorbers
for which D has been detected.
Namely, C has been detected in the
best-measured Lyman limit systems (Burles \& Tytler \cite{bt98a,bt98b};
O'Meara \etal\ \cite{omeara}), and in all three
cases $[{\rm C/H}] \sim -2$,
and  $[{\rm C/O}] \la 0$ or $[{\rm C/Si}] \la 0$.
Results for N are similar.
Thus, the observations apparently demand that no
significant C or N enrichment (over that of $\alpha$-elements)
takes place.

These observations place key constraints on a
putative intermediate-mass Population III.
If the yields of zero-metallicity intermediate mass stars
are comparable to those of their low-metallicity counterparts
(e.g., van den Hoek \& Groenewegen \cite{vdh}), then
indeed, the resulting enrichment is extremely large if
most of the baryons are processed through white dwarfs.
Although the enrichment is reduced when only a fraction of
all baryons are processed, even here, to make a
white dwarf halo at 10\% of the total leads to C and N 
abundances at about $[{\rm C,N/H}] = -1$.  As this violates
the observational constraints, we conclude that
a halo white dwarf population cannot be created at
high redshift if $Z=0$ intermediate-mass stars create
significant C and N.  It then also follows that
the dispersion in D due to chemical evolution is
small; a similar conclusion was reached
by Prantzos \& Ishimaru \cite{pi}.

It is far from clear, however, that zero-metallicity
intermediate-mass stars behave in this way.
In fact, the nucleosynthetic yields for these stars 
are quite uncertain.  In this
mass range, at all metallicities, the
yields depend sensitively on the details
of several ``dredge-up'' events in which convective
mixing brings nuclear burning products up to the
envelope of the star.
These have been studied in detail for solar metallicity
stars and for moderately metal-poor stars, and
the main results of different groups are in 
reasonable qualitative
and quantitative agreement.  However,
there remain basic qualitative uncertainties 
in the post-main sequence evolution of zero-metallicity 
intermediate-mass stars.

Specifically, it is as yet unclear---i.e.,
model-dependent---whether the last and most important of 
these mixing events (the ``third dredge-up'') occurs.
There has been considerable effort recently to 
answer this question, but the studies have given conflicting results.
Chieffi, Dom\'{\i}nguez, Limongi, \& Straniero (2001)
do find such a phase, and conclude that 
zero-metallicity intermediate-mass stars should
be important sources of C and N.
The work of Marigo, Girardi, Chiosi, \& Wood (2001)
does not find a third dredge-up, but the authors are uncertain
whether such a phase would have occurred had they run
their models longer; what {\em is} clear is that the
C and N envelope abundances {\em prior} to this
phase are small ($\la 10^{-3}$ of solar) and so would
not create a significantly signature.  Finally, 
Fujimoto, Ikeda, \& Iben (2000) argue that
for stars with $M \ga 4 \msol$ and $[{\rm Fe/H}] \la -5$,
there is no third-dredge up at all.

It thus appears that the question of C and N abundances in 
$Z=0$ stars is both complex and unsettled.
To simultaneously explain a halo white dwarf population and D depletion
at high redshift
requires that these stars do not eject significant
C and N.  For the moment, this amounts to an assumption,
but one with some support from ongoing studies.

To make the proposed scenario concrete and quantitative,
we have constructed chemical evolution models
which lead to significant D astration at low metallicity
and also have significant white dwarf production.
We will present one such model; of course, given the
freedom to adjust the IMF (and star formation history),
a large class of models are allowed.  
However even with this freedom, the acceptable models
share common traits and make similar predictions.
We have explored the case of a closed-box model.
The stellar creation or birth function $C(m,t)$ 
encodes the rate and mass spectrum of new star
birth via
\beq
{\rm d}N_\star =  C(m,t) \; {\rm d}m \; {\rm d}t
\eeq
We adopt the 
``bimodal'' 
form $C(m,t) = \psi_1(t) \phi_1(m) + \psi_2(t) \phi_2(m)$.
We choose
the star formation rate $\psi_1 = (M_1/\tau_1)e^{-t/\tau_1}$ to be a
short-lived  burst of star formation with $\tau_1 = 100$ Myr
and $M_1 = 0.75 M_B$, with an associated IMF 
which is sharply peaked at intermediate mass stars;
this is the population which destroys deuterium and creates
white dwarfs.  Following Adams \& Laughlin \cite{al},
we take the IMF to be lognormal
\beq
\phi_1(m) \propto m^{-1} \exp[-\ln^2(m/m_c)/2\sigma^2]
\eeq
The model we present here 
has a characteristic mass scale $m_c = 5 \msol$
and a dimensionless width $\sigma = 0.07$ (this is very similar
to a gaussian with mean $\mu = 5 \msol$ 
and width $\sigma_G = \sigma m_c = 0.35 \msol$).
Other parameter choices are possible; for higher
$m_c$, the needed $\sigma$ is narrower.
The other term in the creation function describes
standard, quiescent star formation.
We have $\psi_2 = \lambda g(t) M_{\rm gas}$, 
where $\lambda = 0.3 \; {\rm Gyr}^{-1}$, and
the factor $g(t) = 1-e^{-t/0.5 {\rm Gyr}}$ allows for
a smooth transition from the bursting to 
the quiescent star formation rate.
The IMF $\phi_2$ is a power law with slope 1.35 (Salpeter).

\begin{figure}[htb]
\begin{center}
\epsscale{0.3}
\plotone{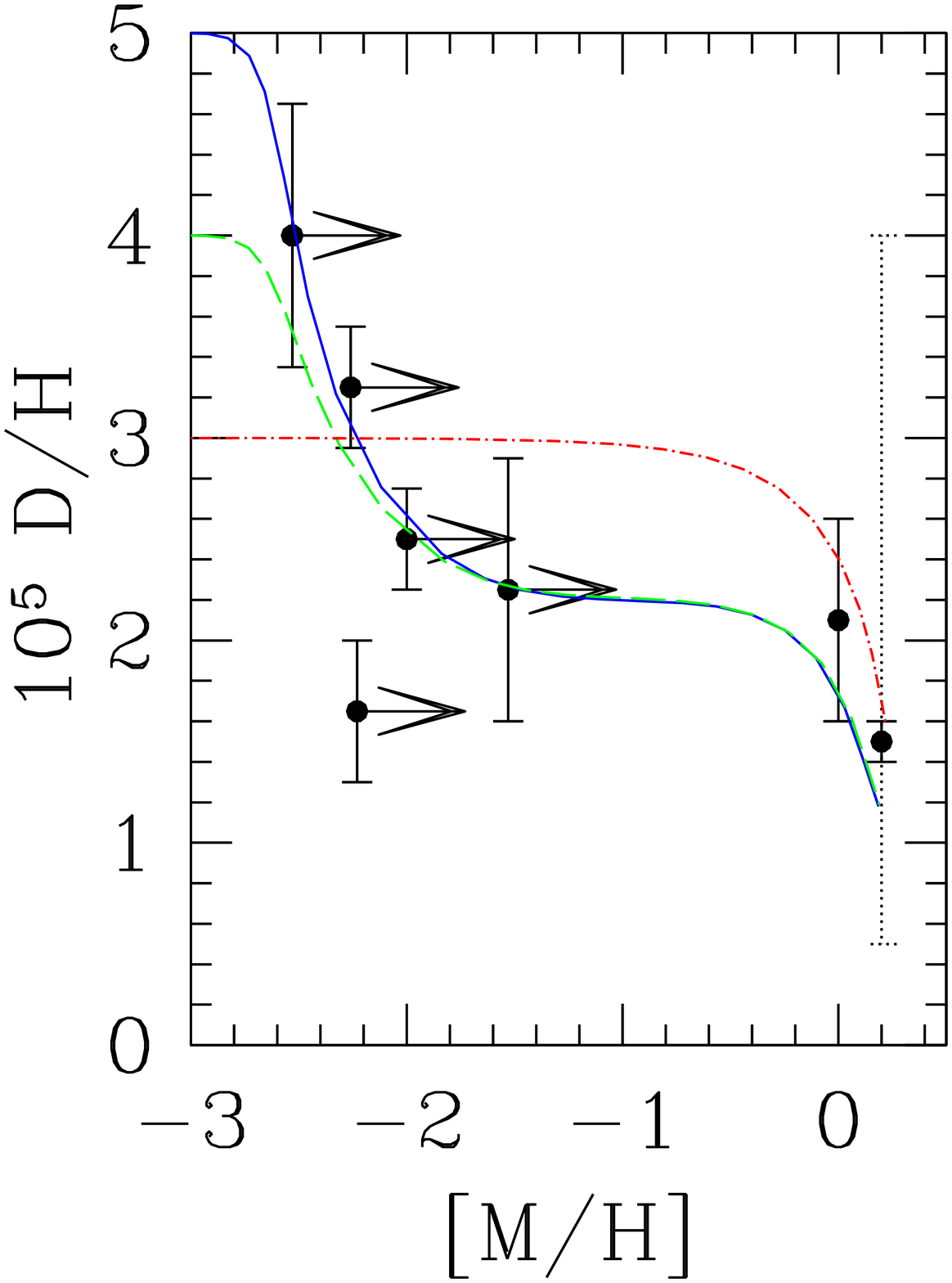}
\plotone{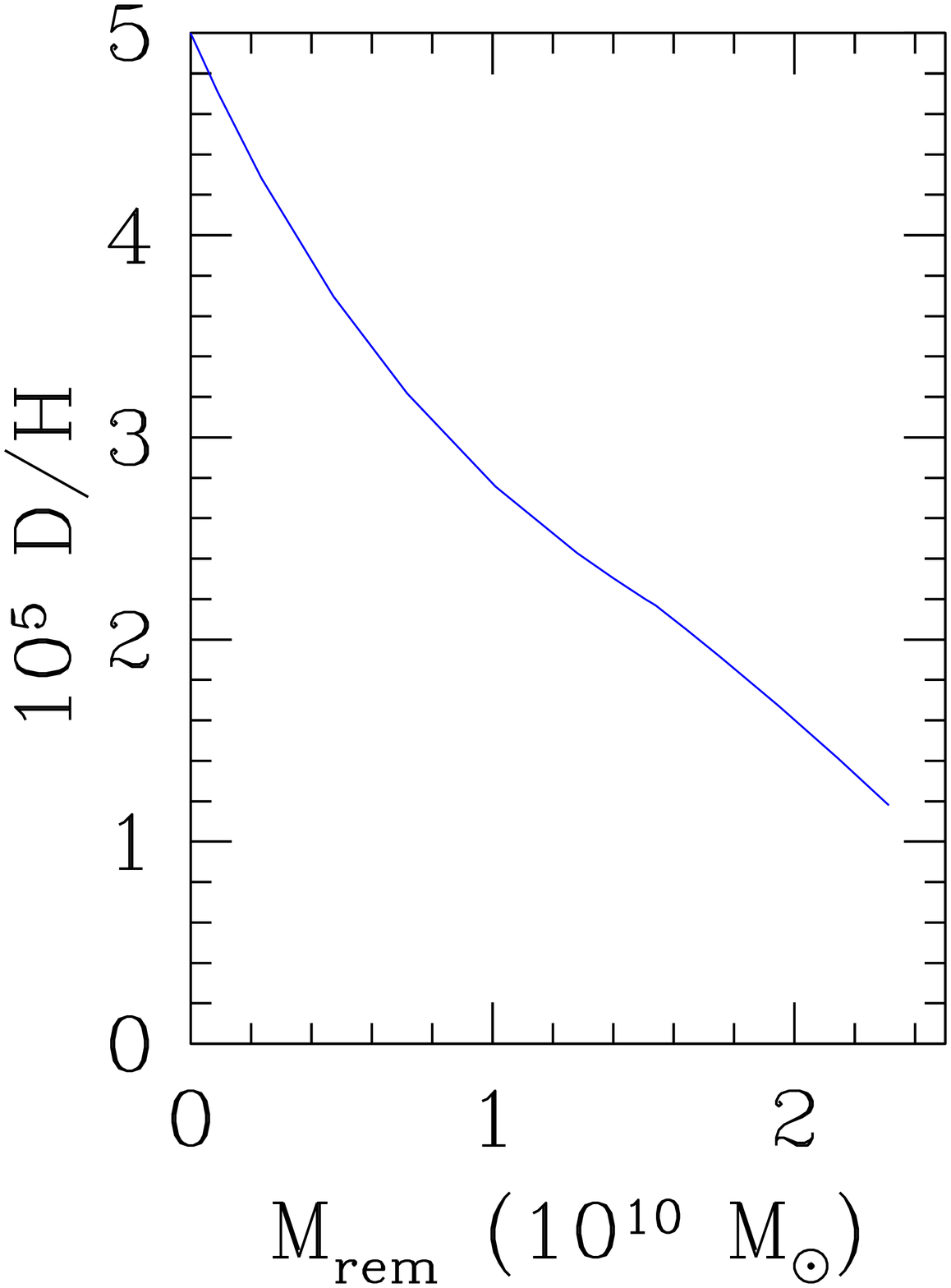}
\plotone{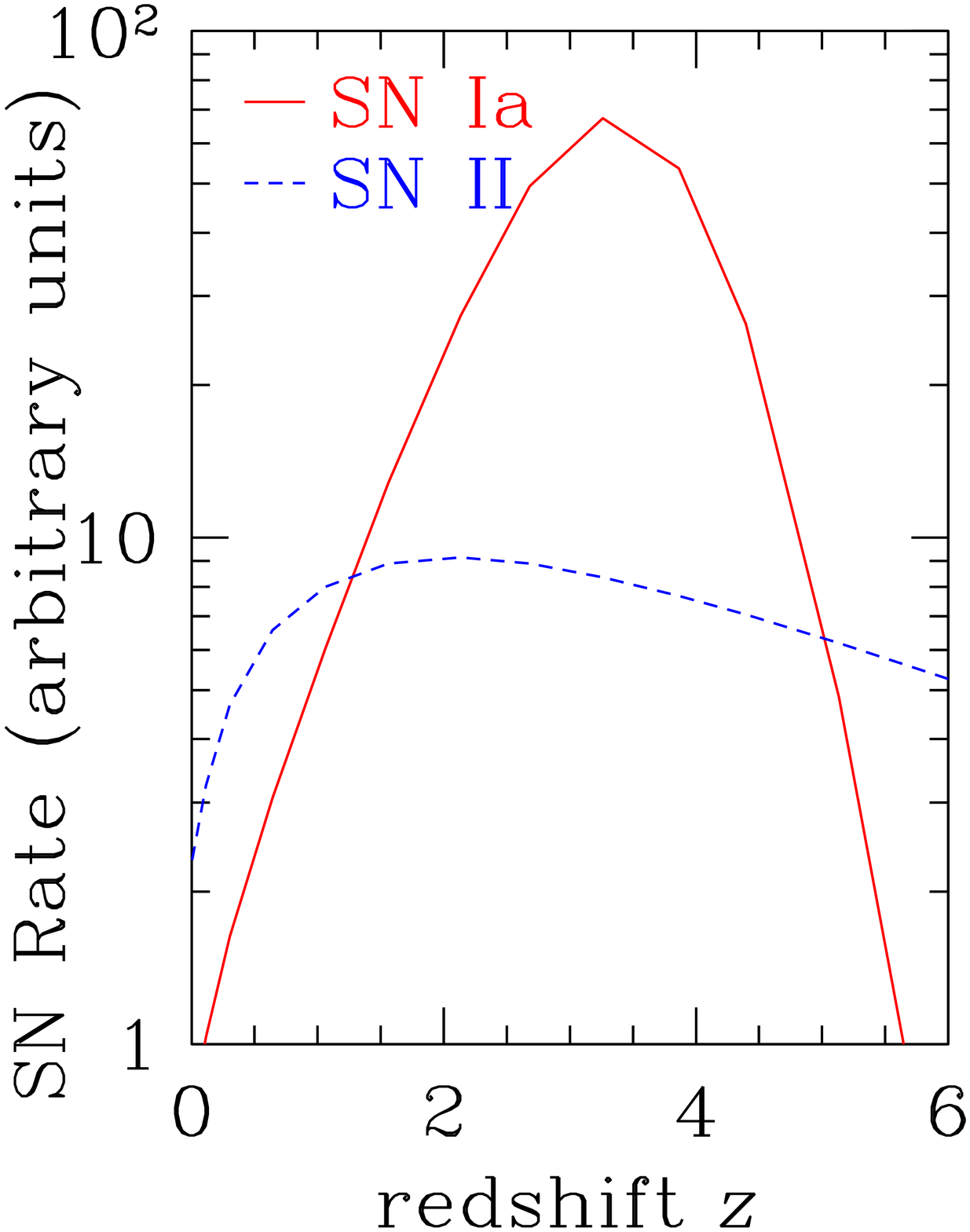}
\end{center}
\caption{Results from the bimodal star formation model.  
{\bf (a)} D evolution as a function of metallicity.
Data and model curves are as noted in the text; the
protosolar value is that of  Geiss \& Gloeckler (1998);
the ISM value and errorbar of Linsky \cite{linsky} appears
as the solid line, while the range reported by Vidal-Madjar
appears as the dotted errorbar.
{\bf (b)} D evolution as a function of remnant production
as measured by the total Galactic mass $M_{\rm rem}$ in remnants,
almost all of which are white dwarfs.
{\bf (c)} The predicted cosmic supernova rates, in arbitrary but
comoving units, as a function of reshift.
\label{fig:bimodal}
}
\end{figure}

Results from the model appear in Figure \ref{fig:bimodal}.
In panel (a) we see the evolution of D/H versus
metallicity [O/H] (as O is an $\alpha$-element which is co-produced with
Si, we take [O/H] = [Si/H]).  Three models are plotted.
The solid curve corresponds to a bimodal model in which
we have taken
a primordial value ${\rm D/H} = 5.0 \times 10^{-5}$.
The two phases of star formation
are clear, as each leads to a characteristic drop in D, 
the burst phase acting at low metallicity and the quiescent
phase acting at high metallicity.
The points are the observed data discussed earlier.
The values from quasar absorption line systems
have an arrow indicating the Si abundances are
lower limits.\footnote{This arises because
the Si abundances are
measured in the gas phase,
and do not include any Si which has been depleted onto dust grains.
The presence of grain depletion complicates attempts to 
interpret measurements and trends in high-redshift abundances,
and to make matters 
worse, the depletion levels found in the Milky Way (more than
an order of magnitude)
are apparently not always appropriate for high-redshift systems.
As shown for the case of DLAs
(e.g., Pettini, Ellison, Stidel, Shapley, \& Bowen 2000;
Molaro \etal\ 2000;
Prochaska \& Wolfe 1999;
Lauroesch, Truran, Welty, \& York)
Si/Zn can be much higher than in the ISM.
Si/Zn is typically about 1/3 (i.e., 0.5 dex), and varies
from system to system, 
sometimes exceeding the solar ratio.}
The dashed curve in Figure 1 corresponds to a bimodal model with
${\rm D/H} = 4.0 \times 10^{-5}$.  In this case, there is
less astration needed at low metallicity, so the amount of
gas processed in the Pop III burst is lower, i.e., $M_1 = 0.5 M_B$ (as is
the  resulting mass of halo white dwarfs).  Note that the
two model curves join once the D has been destroyed in
the Pop III phase, and follow the same quiescent evolution afterwards.
We see that in both cases, the curves provide a reasonable
fit to the data. Therefore, it is difficult to use the
existing D/H data and chemical evolution to accurately predict the
primordial D/H value and hence $\om$. Finally, the dot-dashed curve
corresponds to a model in which there is no Pop III and a primordial value
${\rm D/H} = 3.0 \times 10^{-5}$ as in O'Meara et al.\ \cite{omeara}.
In this ``standard'' case there is essentially no D depletion at
low metallicity and, as we have noted, the dispersion in D is
unexplained.

The deaths of the Pop III intermediate-mass stars not only 
release D-free material, but also produce white dwarfs.
This correlation is illustrated in 
Figure \ref{fig:bimodal}(b); we plot the 
D abundance versus the total mass in compact remnants,
essentially all of which is in white dwarfs.  
The model used here is that shown in the solid curve of
Figure 1(a).
We see that the burst phase, during which D/H drops to 
about $2.1 \times 10^{-5}$, puts about
15\% of the baryons (about $1.5 \times 10^{10} \msol$) into
white dwarfs.  The quiescent phase adds another $10^{10} \msol$.
Note well the distinction between this work and previous work.
Previous work was centered on the possibility of a white dwarf halo.
Here we are interested in D/H and producing a few white dwarfs. A halo
population of a few percent would dramatically ease the previous constraints.
Indeed, Figure 1(b) shows 
that the 2.5\% observation of Oppenheimer is more than sufficient.

Note that by showing the extragalactic D {\em vs}.\ Si data on the same
plot as the Galactic points, we implicitly assume that
the Galaxy is a system representative of the quasar absorbers.
It then would also follow that the white dwarf and carbon
production also would apply to these systems.
However, one need not take the model so literally as to
demand that all baryonic matter--or even 
all the baryons in protogalaxies--follow this evolution.
For example, in a hierarchical merging picture, one could
instead imagine that this evolution is common to
some but not all of the subsystems
which form the Galaxy.  Thus, D depletion (and 
carbon and white dwarf production) would be inhomogeneous
within the protogalaxy, and across the observed quasar absorbers.

Figure \ref{fig:bimodal}(c) shows 
the predicted supernova rates, both for
Type II as well as Type Ia events.
The model use here is that shown in the solid curve of
Figure 1(a) and in Figure 2.
The Type II rate follows the assumed star formation
rate, which has simply been taken to be 
$\psi \propto M_{\rm gas}$ and which turns on sequentially 
after the bursting Pop III star formation ends.
The Type Ia rate is calculated using the prescription of
Matteucci \& Greggio (1986), and assumes that the
binarity properties of the Pop III stars
are the same as those observed locally today.  The Type Ia rate
peaks sharply, echoing the rapid burst and small
mass range that characterized the Pop III star formation.
Note that while the peaked nature of the Ia rate
is a generic feature of the model we propose, the
peak redshift is model-dependent, reflecting the
characterist mass (and thus the lifetime) of the Pop III
stars.  Figure 1(c) shows the results for a value
$m_c = 5\msol$, which gives a Type Ia peak at $z \sim 3$.
Had we chosen $m_c = 6\msol$ ($4\msol$), we would find a peak at
$z \sim 5$ (2).
In any case, we predict that the Type Ia supernova rate
is quite large at moderate redshift; this
can be tested in ongoing high-$z$ supernova searches.
Note that a suppression 
of Type Ia supernovae at low metallicities (Kobayashi \etal \cite{koba}),
if real, could  alter this prediction.

A potential constraint on high-redshift
Type Ia supernova activity also arises from the
cosmic $\gamma$-ray background
at MeV range.  The dominant source of photons at these energies
is mostly nuclear line 
emission due to the 
$\iso{Ni}{56} \rightarrow \iso{Co}{56} \rightarrow \iso{Fe}{56}$
decay chain
(e.g., Ruiz-Lapuente, Cass\'e, \& Vangioni-Flam \cite{rvc}).
The enhanced SN Ia production of our model 
increases the MeV background relative to the
predictions expected from a constant-IMF cosmic star formation
rate.  However, most of the contribution is at high redshift and
is thereby diluted.  In fact, the background is not
a good discriminant of these epochs, and the
Type Ia rate proposed here is non inconsistent with
the cosmic MeV background.

\section{Conclusions}

The present work is motivated by the reports of
a population of halo white dwarfs in
our Galaxy, and of 
dispersion in the high redshift D/H values.
In light of these recent observations, we have explored the 
consequences of a high redshift population of 
intermediate mass stars.  This population leads
to D destruction at low metallicity, as well as
white dwarf production, and would add to the
far-IR background without increasing the local
near-IR background.  For this scenario to remain viable,
it is necessary that the C and N yields of
zero-metallicity stars are small, as suggested
by some recent stellar evolution models.
We have constructed relatively simple models which fit the data
reasonably well for primordial abundances of D/H =  4 and 5 $\times
10^{-5}$. Other values are possible by merely adjusting the astration
constant,
$M_1$, in the POP III star formation rate.  Thus, we conclude that it is
difficult to use existing data to accurately predict either the
primordial abundance of D/H or $\om$.

An interesting prediction of a halo in which the white dwarf number 
exceeds that in the disk by an order of magnitude
is that, binarity fractions being comparable, the number of 
Type Ia would
be greatly boosted in the very early universe.
Some may even be seen in the halos of nearby galaxies,
(Smecker \& Wyse 1991), though the spatial distribution of these
events within galaxies will depend on the interplay between the formation
and evolution of the Pop III stars and protogalaxies.

A potentially disturbing complication is
raised by the work of Hansen (2001), who estimates
the white dwarf ages from their luminosity and
color distribution.
He finds that they are relatively young and
if true, this is inconsistent with our proposal.
On the other hand, if the white dwarfs are real
and in the halo, a young population is even more
difficult to explain than an old one.
Given this, and the difficulty
of the age calculations (which also couple to
the difficulty in determining yields
accurately), 
we require that the Pop III white dwarf relics are old.
However,
further study of this issue is clearly warranted.

The recently reported observations
of an intrinsic  dispersion in high-redshift D,
taken together with the observation of a new Galactic white dwarf population,
demand a non-standard history of star formation and chemical
evolution.  If these observations persist, then an unconventional scenario
of the kind we propose will be {\em required}.
We note that standard scenarios of chemical
evolution are unable to lead to significant dispersion
at low metallicity. If these standard models prevail, then one must
concur  with Pettini \& Bowen (2001) that 
the D dispersion at high redshift is an artifact,
and that significant systematic errors are present.
At present, the use of D/H as a precision baryometer must be called
into question until these systematic errors are
better understood. We expect that future analyses and observations will
either show a convergence towards a unique observed 
  D/H value, representative of the primordial one, or reveal the
 dependence of D/H on metallicity in high z absorbing clouds.

\acknowledgements
It is a pleasure to thank Robert Mochkovitch, Patrick Petitjean and
  Jim Truran
for stimulating discussions.
BDF acknowledges enlightening conversations
with the participants of the ``Deuterium in the Universe''
workshop.
The work of KAO was supported in part by DOE grant
DE--FG02--94ER--40823 at the University of Minnesota.
KAO also acknowledges support during a sabbatical at CERN where much of
this work was done.
This work has been supported in part by PICS 1076 from the CNRS.

\end{document}